\documentclass[aps,pre,reprint,superscriptaddress]{revtex4-1}
\pdfoutput=1

\usepackage{amsmath}
\usepackage{graphicx}
\usepackage{hyperref}
\usepackage{color}

\graphicspath {{images/}}

\newcommand{\ly}{\l_Y}

\newcommand{\ku}{k_\uparrow}

\newcommand{\kd}{k_\downarrow}

\newcommand{\kp}{k_+}

\newcommand{\km}{k_-}

\newcommand{\dg}{\Delta G}

\newcommand{\dir}[2]{\frac{\mathrm{d}#1}{\mathrm{d}#2}}

\begin{document}

\author{Rory A. Brittain}
\author{Nick S. Jones}
\affiliation{Department of Mathematics, Imperial College London, London, SW7 2AZ, UK}
\author{Thomas E. Ouldridge}
\email{t.ouldridge@imperial.ac.uk}
\affiliation{Department of Bioengineering, Imperial College London, London, SW7 2AZ, UK}

\title{What we learn from the learning rate\thanks{This is an author-created, un-copyedited version of an article published in Journal of Statistical Mechanics: Theory and Experiment.  IOP Publishing Ltd is not responsible for any errors or omissions in this version of the manuscript or any version derived from it. The Version of Record is available online at \url{https://doi.org/10.1088/1742-5468/aa71d4}.}}

\begin{abstract}
The learning rate is an information-theoretical quantity for bipartite Markov chains describing two coupled subsystems. It is defined as the rate at which transitions in the downstream subsystem tend to increase the mutual information between the two subsystems, and is bounded by the dissipation arising from these transitions. Its physical interpretation, however, is unclear, although it has been used as a metric for the sensing performance of the downstream subsystem. In this paper, we explore the behaviour of the learning rate for a number of simple model systems, establishing when and how its behaviour is distinct from the instantaneous mutual information between subsystems. In the simplest case, the two are almost equivalent. In more complex steady-state systems, the mutual information and the learning rate behave qualitatively distinctly, with the learning rate clearly now reflecting the rate at which the downstream system must update its information in response to changes in the upstream system. It is not clear whether this quantity is the most natural measure for sensor performance, and, indeed, we provide an example in which optimising the learning rate over a region of parameter space of the downstream system yields an apparently sub-optimal sensor.
\end{abstract}

\maketitle

\section{Introduction}
The mathematical theory of communication was founded by Claude Shannon in 1948~\cite{Shannon_A_1948}. He was concerned with how to transfer a message from one point to another, and described the input signal through a random variable $X$ and the output by a random variable $Y.$
He introduced the Shannon entropy $H[X],$ which quantifies the {\it a priori} uncertainty in $X,$
\begin{equation}
	H[X]=-\sum_xp(x)\ln p(x),
\end{equation}
where $x$ labels the possible outcomes of the (discrete) random variable $X.$ The logarithm can be taken with respect to any base; we choose to use natural logarithms, in which case entropy is measured in nats. If a meaningful signal is passed between $X$ and $Y,$ then knowledge of $Y$ should reduce the uncertainty in $X.$ The average uncertainty in $X$ that remains given knowledge of $Y$ is given by the conditional entropy
\begin{equation}
	H[X|Y]=-\sum_{xy}p(x,y)\ln p(x|y),
\end{equation}
where $y$ labels the possible outcomes of $Y.$ It is necessarily true that $H[X|Y] \leq H[X]$; it is not possible to become more uncertain about $X$ by knowing $Y$~\cite{Elements_of_Information_Theory}. The difference between the two quantities is  the mutual information between $X$ and $Y$ \cite{Elements_of_Information_Theory},
\begin{equation}
	I[X;Y]=H[X]-H[X|Y]=\sum_{xy}p(x,y)\log\frac{p(x,y)}{p(x)p(y)}.
\end{equation}
The mutual information $I[X;Y]$ is symmetric with respect to switching $X$ and $Y.$

The terminology ``entropy" is suggestive of a link with thermodynamics. In fact, the entropy $H[X]$ is precisely the generalised non-equilibrium entropy of a thermodynamic system described by $X$ \cite{crooks_entropy_1999,Jarzynski2011,Seifert2012}. Information is therefore indicative of a low entropy state, and recent work has focussed on the thermodynamic consequences of generating information \cite{Sagawa2009, Sagawa2012, still_thermodynamics_2012, horowitz_thermodynamics_2014, hartich_stochastic_2014,  parrondo_thermodynamics_2015, Bo_Thermodynamic_2015, ouldridge_thermodynamics_2017, Ouldridge_polymer_2016}, and the possibility of exploiting information to do useful work \cite{Bauer2012, Horowitz2013, Chapman2015, McGrath_A_2017, Yamamoto_Linear_2016}. This body 
of work provides an explanation of Maxwell's infamous demon \cite{Bennett2003} that does not rely on  ``erasure" of memories \cite{Landauer1961,Bennett2003}. The demon, by measuring its environment, appears to be able to extract work from equilibrium fluctuations. However,
the demon can only extract work from its measurements if they generate non-equilibrium mutual information, and generating this mutual information in the first place requires consumption of thermodynamic resources \cite{parrondo_thermodynamics_2015, ouldridge_thermodynamics_2017}.

A key area in the interplay between information and thermodynamics is in biochemical sensing. Given that cells have a limited supply of resources, it is reasonable to ask whether these are put to optimally efficient use in sensing their environment --- and if not, to explain why. Knowing the fundamental thermodynamic limits on sensor operation is also relevant to the engineering  of synthetic sensors. Drawing on the connections between information and thermodynamics, several groups have considered the intrinsic costs associated with sensing and adaption (sensors that adapt revert to zero after lengthy exposure to a constant input)   \cite{Tu2008, Lan2012, Sartori_Thermodynamic_2014,
Mehta_Energetic_2012, govern2014, Govern_PRL_2014,
tenWolde_Fundamental_2016,
Mancini_Trade-Offs_2016,
ito_maxwells_2015,ouldridge_thermodynamics_2017,Bo_Thermodynamic_2015, Das_A_2016}. For example, Govern and ten Wolde showed that molecular readout molecules of cell-surface receptors must consume free energy to form long-lived memories of receptor states \cite{Govern_PRL_2014, govern2014}, a process equivalent to thermodynamic measurement of the kind performed by Maxwell's demon~\cite{ouldridge_thermodynamics_2017}. Similarly, Bo, Del Giudice and Celani  showed that the information  stored by the molecular readouts about the entire trajectory of the receptors is bounded by the dissipation of the system minus the dissipation of the receptor transitions alone \cite{Bo_Thermodynamic_2015}.

A quantity, $\ly,$ called the learning rate has been proposed as a metric for the performance a sensor when the signal/sensor system is modelled as a bipartite Markov chain \cite{barato_efficiency_2014, hartich_sensory_2016}. The learning rate is bounded by the entropy production of the sensor so this appears to put an energetic bound on sensing quality. However, it is not clear that it is the most natural measure of sensing.

In this paper, we will consider the behaviour of the learning rate for three simple steady-state systems, identifying and explaining differences between the physical content of $\ly$ and $I[X;Y].$ Typically,  differences in behaviour arise because the learning rate quantifies the rate at which transitions in $Y$ must act to restore information between $X$ and $Y,$ which is not necessarily closely related to the steady-state level of information or correlation. As a result, we are able to identify a fourth system in which optimizing over a parameter of the sensor at fixed input signal dynamics gives markedly different results when using the learning rate and the mutual information as metrics. In this case, optimising the mutual information provides the more intuitively reasonable optimal sensor.

\section{Set up: Sensing and thermodynamics in bipartite Markov chains}
Bipartite Markov chains \cite{horowitz_thermodynamics_2014, hartich_stochastic_2014} are a central tool in modelling Maxwell's demon-like behaviour and cellular sensing circuits. A bipartite Markov chain has states that are labelled with two variables $x$ and $y,$ which are generally taken to correspond to the states of two subsystems $X$ and $Y.$ Transitions only change one of the variables so the transition rate from $(x,y)$ to $(x',y')$ is
\begin{equation}
	w_{yy'}^{xx'}=\left\{
	\begin{array}{lr}
		w_y^{xx'} & \text{if } y=y' \text{ and } x\neq x',\\
		w_{yy'}^x & \text{if } y\neq y' \text{ and } x=x',\\
		0 & \text{if } y\neq y' \text{ and } x\neq x'.
	\end{array}
	\right.
\end{equation}
These rates have dimensions of inverse time. Throughout this paper we will use arbitrary units. The systems we will consider have a discrete state space and continuous time.

The second law of thermodynamics constrains the entropy production of the system to be positive. We can use the bipartite structure to separate the entropy production, $\dot{S}_i$, into six parts \cite{horowitz_thermodynamics_2014}
\begin{equation}
	\dot{S}_i=\mathrm{d}_tS^X+\dot{S}_r^X-\dot{I}^X+\mathrm{d}_tS^Y+\dot{S}_r^Y-\dot{I}^Y\geq 0
	\label{eq:entropyproduction}
\end{equation}
where the rates of increase of entropy of the $X$ and $Y$ subsystems are
\begin{align}
	\mathrm{d}_t S^X&=\sum_{xx'y}w_y^{xx'}p(x,y)\log\frac{p(x)}{p(x')},\nonumber\\
	\mathrm{d}_t S^Y&=\sum_{xyy'}w_{yy'}^xp(x,y)\log\frac{p(y)}{p(y')},
\end{align}
the rates of increase of the entropy of the environment due to transitions in $X$ and $Y$ are
\begin{align}
	\dot{S}_r^X&=\sum_{xx'y}w_y^{xx'}p(x,y)\log\frac{w_{y}^{xx'}}{w_y^{x'x}},\nonumber\\
	\dot{S}_r^Y&=\sum_{xyy'}w_{yy'}^xp(x,y)\log\frac{w_{yy'}^x}{w_{y'y}^{x}}
\label{eq:Sr}
\end{align}
and
\begin{align}
	\dot{I}^X&\equiv\sum_{xx'y}w^{xx'}_yp(x,y)\log\frac{p(y|x')}{p(y|x)},\nonumber\\
	\dot{I}^Y&\equiv\sum_{xyy'}w^x_{yy'}p(x,y)\log\frac{p(x|y')}{p(x|y)}.
	\label{eq:informationflows}
\end{align}
Note that in Equation \ref{eq:Sr}, if a single change of state of $X,Y$ can be associated with multiple distinct reaction pathways, these separate pathways must be explicitly considered as separate contributions to the sum \cite{Yamamoto_Linear_2016}.
The flows $\dot{I}^X$ and $\dot{I}^Y$ are so called because they represent changes in the mutual information between $X$ and $Y$ due to transitions in $X$ and $Y$ respectively,
\begin{equation}
	\dir{}{t}I[X;Y]=\dot{I}^X+\dot{I}^Y.
\end{equation}

Importantly, the second law holds not just for Equation~\ref{eq:entropyproduction}, but also for each subsytem  individually \cite{allahverdyan_thermodynamic_2009,horowitz_thermodynamics_2014,barato_efficiency_2014}
\begin{align}
	\mathrm{d}_t S^X+\dot{S}_r^X-\dot{I}^X&\geq0,\nonumber\\
	\mathrm{d}_t S^Y+\dot{S}_r^Y-\dot{I}^Y&\geq0.
	\label{eq:bipartite_second_law}
\end{align}
This form of the second law has been extended by Horowitz \cite{Horowitz_Multipartite_2015} to systems with multiple subsystems that have independent noise, and has been used by Goldt and Seifert \cite{Gold_Stochastic_2017} to find the thermodynamic bound on the information learned by a feedforward neural network with Markovian dynamics.

We can identify $\sigma^X=\mathrm{d}_t S^X+\dot{S}_r^X$ and $\sigma^Y=\mathrm{d}_t S^Y+\dot{S}_r^Y$ as the entropy productions of $X$ and $Y$, ignoring the correlations between the two subsystems. $\dot{I}^Y < 0$ then allows for $\sigma^Y < 0$ --- correlations between subsystems allow for apparent violation of the second law if the subsystems are erroneously considered in isolation. If ${\sigma}^Y < 0 $ without direct energy flow from $X$ to $Y,$ $X$ can be interpreted as a demon acting on subsystem $Y.$ In terms of sensing, $Y$ is generally interpreted as a sensory network for the signal $X.$

In steady state the mutual information is constant and so
\begin{equation}
	\dot{I}^X+\dot{I}^Y=0.
	\label{eq:steadystate}
\end{equation}
In this case, transitions in one subsystem must reduce the mutual information as much as the transitions in the other subsystem increase it.

The quantities $\dot{I}^Y$ and $\dot{I}^X$ have been introduced in Equation~\ref{eq:entropyproduction} to separate the second law into two parts. But since $\dot{I}^Y$ represents the rate at which transitions in $Y$ tend to increase information between $X$ and $Y,$ it has been named the learning rate $\ly = \dot{I}^Y$  and proposed as a metric for the performance of $Y$ as a sensor of $X$ \cite{barato_efficiency_2014, hartich_sensory_2016}. Identifying $\ly$  as a measure of sensory performance also sets a thermodynamic bound on sensor function, since from Equation~\ref{eq:bipartite_second_law}, $\ly \leq \sigma^Y,$ the entropy generation due to transitions in $Y.$ It is straightforward to show that one can equivalently write the learning rate as \cite{hartich_sensory_2016}
\begin{equation}
	\ly=\dir{}{\tau}I[X_t;Y_{t+\tau}]\Big|_{\tau=0}.
\label{eq:dIdt}
\end{equation}
Intuitively, $\ly$ is the degree to which future values of $Y$ are more predictive of the current value of $X$ the current value of $Y$ is, due to $Y$ continuing to learn about $X.$ 

 Although the use of the learning rate to quantify sensor performance seems reasonable from its definition, its physical interpretation has not been extensively explored. Other authors, such as  Das {\it et al.} \cite{Das_A_2016}, have preferred the instantaneous mutual information, $I[X;Y],$ over $\ly$ as a metric for  performance when deriving thermodynamic constraints on sensor function. Is $\ly$ essentially reporting the degree of interdependence between $X$ and $Y,$ similar to quantities like $I[X;Y]$ or the covariance $\mathrm{Cov}(X,Y)$? Or does it highlight distinct features of the network? Is $\ly$ informative when trying to build optimal sensors for a given input signal? A further interpretational challenge occurs in steady state (a common setting for sensors). In this case \cite{barato_efficiency_2014, hartich_sensory_2016, Das_A_2016}
\begin{equation}
	\ly=-\sum_{xyy'}w^x_{yy'}p(x,y)\log\frac{p(x,y)}{p(x,y')},
	\label{eq:other_defn}
\end{equation}
and
\begin{equation}
	\ly= -\dir{}{\tau}I[X_{t+\tau};Y_t]\Big|_{\tau=0}.
	\label{eq:nostalgia}
\end{equation}
Equation~\ref{eq:nostalgia} suggests that learning rate also reflects ``nostalgia", a quantity defined for discrete-time systems 
by Still {\it et al.}  \cite{still_thermodynamics_2012}. In steady state, $\ly$ not only reflects how much more the sensor $Y$ learns about the current value of $X$ as time progresses incrementally (Equation~\ref{eq:dIdt}), but also the degree to which $Y$ has nostalgic information about the history of $X$ that is irrelevant to future signals (Equation~\ref{eq:nostalgia}). From this ``nostalgia" perspective, it is unclear why large values of $\ly$ would correspond to effective sensing.

\section{Results and Discussion}
\subsection{The learning rate reports on correlations in the simplest bipartite sensing system}
\label{sec:system1}
\begin{figure}
	\centering
	\includegraphics[width=\linewidth,trim={0 4.4cm 0 4.6cm},clip]{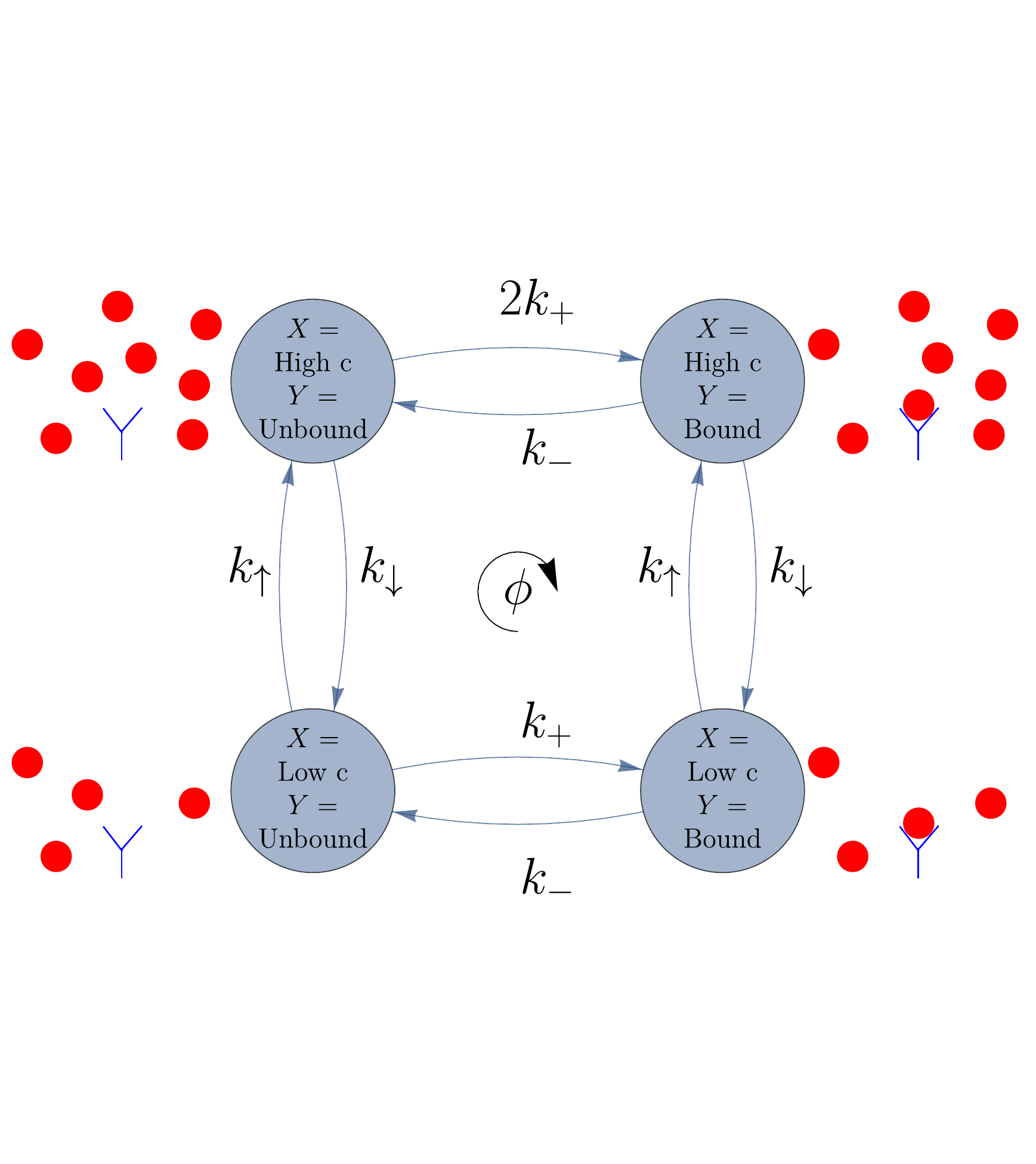}
	\caption{The states and transitions for a simple system involving a two-state signal and a two-state sensor, in which the signal transitions are independent of the sensor state. We illustrate a biophysical interpretation of such a system, corresponding to an environment that can switch between high and low concentrations of ligand molecules, and a sensor that can bind to and unbind from the ligands. The rate of binding is proportional to the ligand concentration and the high concentration state has twice the concentration of the low concentration state in this illustration. The ligand concentration is absorbed into the transition rates $\kp$ and $2\kp.$ We illustrate the flux $\phi,$ the net flow of trajectories around the state space in steady state.}
	\label{fig:specialcasegraph}
\end{figure}

We first consider a canonical steady-state sensing system involving a two-state signal $X$ and a two-state sensor $Y,$ in which transition propensities for $X,$ $w^{xx\prime}_y,$ are $Y$-independent  \cite{barato_information_2013,barato_efficiency_2014}. This $2\times2$ system is the simplest possible sensing setup.  In biophysical terms, this system (illustrated in Figure~\ref{fig:specialcasegraph}) would correspond to a single receptor that can bind to ligand molecules, which is exposed to two discrete concentrations of ligands in its external environment \cite{barato_efficiency_2014}. The states $x=0,1$ correspond to low and high concentrations, and  $y=0,1$ to unbound and bound receptors, respectively.

Let us, for convenience, assume that the high concentration is twice that of the low concentration. The rate at which the environment transitions from the low state to the high is $\ku$ and the reverse transition occurs at a rate $\kd,$ independent of the receptor state. We only consider a single receptor molecule, which can be in two states: bound and unbound. The rate for a ligand molecule to bind to the receptor is proportional to the ligand concentration (i.e. mass-action kinetics and a well-mixed system are assumed). The concentrations are absorbed into the transition rates $\kp$ and $2\kp$ for ligand-binding in low and high concentration conditions, respectively. The rate of unbinding, $\km,$ is independent of the ligand concentration. 

\begin{figure*}
	\centering
	\includegraphics[width=0.6\linewidth]{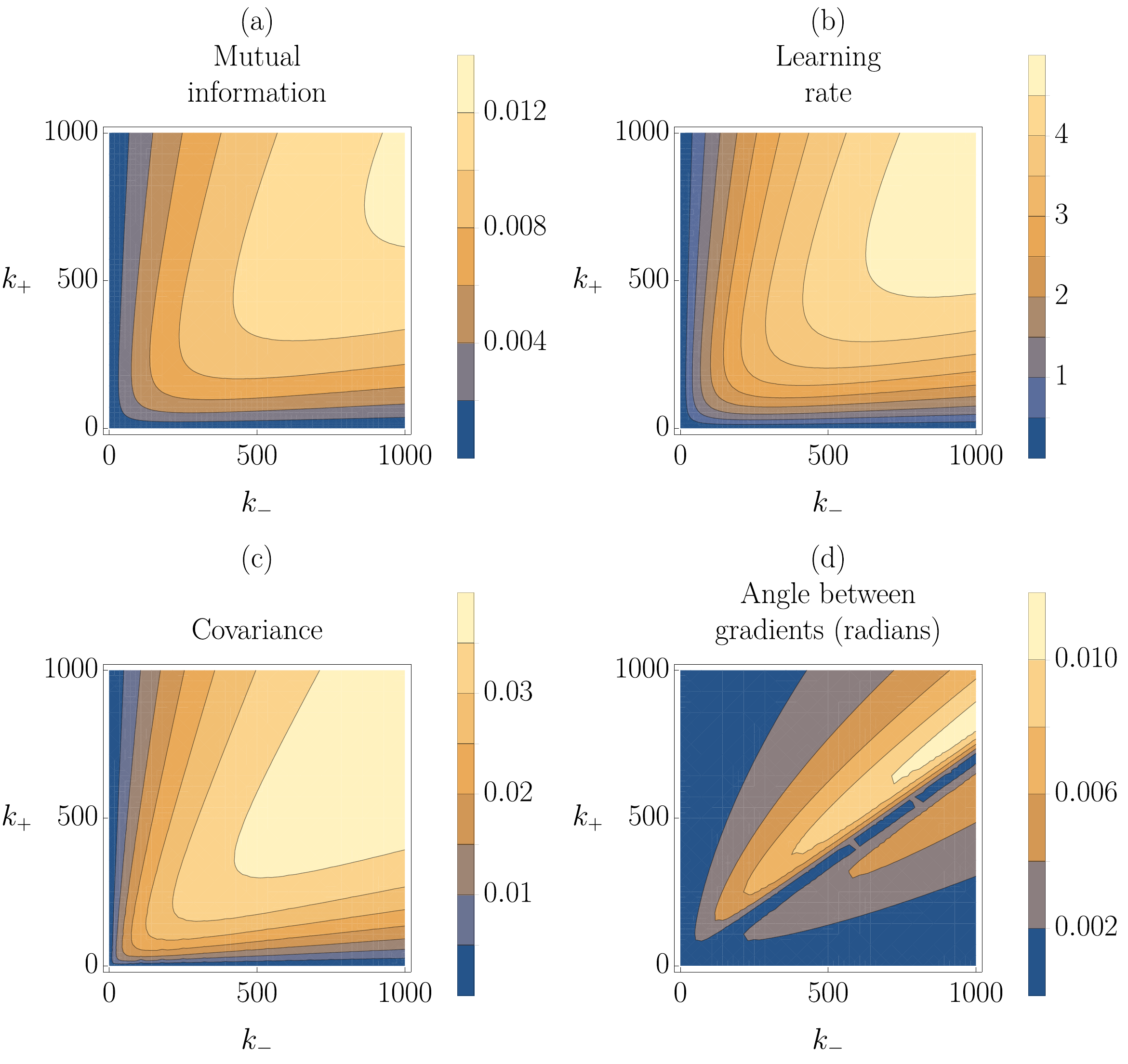}
	\caption{The  mutual information, learning rate and covariance all show similar behaviour as a function of sensor parameters for the simplest possible sensing device, illustrated in Figure~\ref{fig:specialcasegraph}. These quantities are plotted in (a), (b) and (c), respectively, as a function of $\kp$ and $\km,$ for $\ku=\kd=100.$ In (d), we plot the angle between $\nabla I[X;Y]$ and $\nabla \ly,$ where $\nabla$  is defined as differentiation with respect to $(\kp, \km).$The small angles observed suggest that there is little difference between optimizing for $\ly$ and for $I[X;Y].$}
	\label{fig:specialcaseplot}
\end{figure*}


In Figure~\ref{fig:specialcaseplot}, we plot the instantaneous mutual information $I[X;Y],$ the learning rate $\ly$ and the covariance $\rm{Cov}(X,Y)$ as a function of $\kp$ and $\km,$ at fixed $\ku$ and $\kd$ (in this case, $\ku=\kd$). Although the functions are not identical, the dependence on $\kp$ and $\km$ is very similar for all three. In particular, if $I[X;Y]$ is greater at $(\kp,\km)$ than at $(k^\prime_+,k^\prime_-)$ then it is usually true that $\ly$ is greater at $(\kp,\km)$ than at $(k^\prime_+,k^\prime_-).$ The metrics are therefore almost equivalent for the purpose of optimising this sensor design. Deviations from this equivalence are manifest as non-parallel gradients of $I[X;Y]$ and $\ly$ with respect to $\kp$ and $\km$; we plot the angle between gradients as a function of $\kp$ and $\km$ in Figure~\ref{fig:specialcaseplot}\,(d), showing that it is small.

It is possible to understand this similarity from the underlying definition of the learning rate. Using the notation $p(x,y)=p_{xy},$ then Equation~\ref{eq:other_defn} can be expanded as
\begin{align}
\ly=&-\left(( w^0_{01}p_{00} - w^0_{10} p_{01} ) \ln \frac{p_{00}}{p_{01}}\right) \nonumber
\\
&-\left( ( w^1_{10}p_{11} - w^1_{01} p_{10} ) \ln \frac{p_{11}}{p_{10}} \right).
\end{align}
This simple system has one loop of states. In  the steady state, any non-zero clockwise flux $\phi = w^0_{10} p_{01} -  w^0_{01}p_{00}$ between the $(0,1)$ and $(0,0)$ states must be balanced by an equal  clockwise flux for all other pairs of states around the loop, as  illustrated in Figure~\ref{fig:specialcasegraph}.  In particular, $\phi = w^1_{01} p_{10} -  w^1_{10}p_{11}$ so
 \begin{align}
\ly= \phi \ln \frac{p_{00}p_{11}}{p_{01}p_{10}}.
\label{eq:ly}
\end{align}
The logarithm in Equation~\ref{eq:ly} is known as the `information affinity' \cite{horowitz_thermodynamics_2014,Yamamoto_Linear_2016}, and it reflects the informational driving force exerted by one subsystem on the other. Clearly, it is related to the correlation between $X$ and $Y.$ If $X$ and $Y$ are more likely to be in the same state than different states then the affinity is positive. If they are more likely to be in different states then the affinity is negative.  

The flux $\phi$ is the conjugate current to the information affinity \cite{horowitz_thermodynamics_2014,Yamamoto_Linear_2016}. For this system the flux can be written
\begin{align}
	\phi&=p_{11}\kd-p_{01}\ku \nonumber \\
	&=p(Y=1|X=1)p(X=1)\kd \nonumber \\
	&-p(Y=1|X=0)p(X=0)\ku.
\end{align}
In this simple case we have assumed that $X$ transitions are independent of $Y,$ and thus $p(x)$ is determined solely by $\ku$ and $\kd$: $p(X=1) = \ku /(\ku+ \kd)$ and $p(X=0) = \kd /(\ku+ \kd).$ Thus
\begin{equation}
	\phi=\frac{\ku \kd}{\ku+\kd}\Big(p(Y=1|X=1)-p(Y=1|X=0)\Big).
\label{eq:phi}
\end{equation}
The quotient in Equation~\ref{eq:phi} is independent of $\kp$ and $\km.$ It reflects the overall timescale of the process, which is manifest in the learning rate since $\ly$ is a dimensional quantity, and the uncertainty or entropy in $X.$ The second part is clearly related, again, to the correlation between $X$ and $Y$: it is large and positive when they are correlated, and large and negative when anticorrelated, just like the information affinity.

Thus, for a $2\times2$ bipartite system, in which the value of $Y$ does not influence the transitions in $X,$ the learning rate $\ly$ essentially reports on the correlation between $X$ and $Y,$ and is thus closely related to the mutual information $I[X;Y]$ or covariance between $X$ and $Y.$
$\ly$ also incorporates a prefactor that reflects the overall timescale of the process. In Sections~\ref{sec:system2} and \ref{sec:system3}, we shall violate the assumptions that lead to this conclusion and explore the consequences for the relationship between learning rate and mutual information.

\subsection{Feedback from the downstream to the upstream system}
\label{sec:system2}
\begin{figure}
	\centering
	\includegraphics[width=\linewidth]{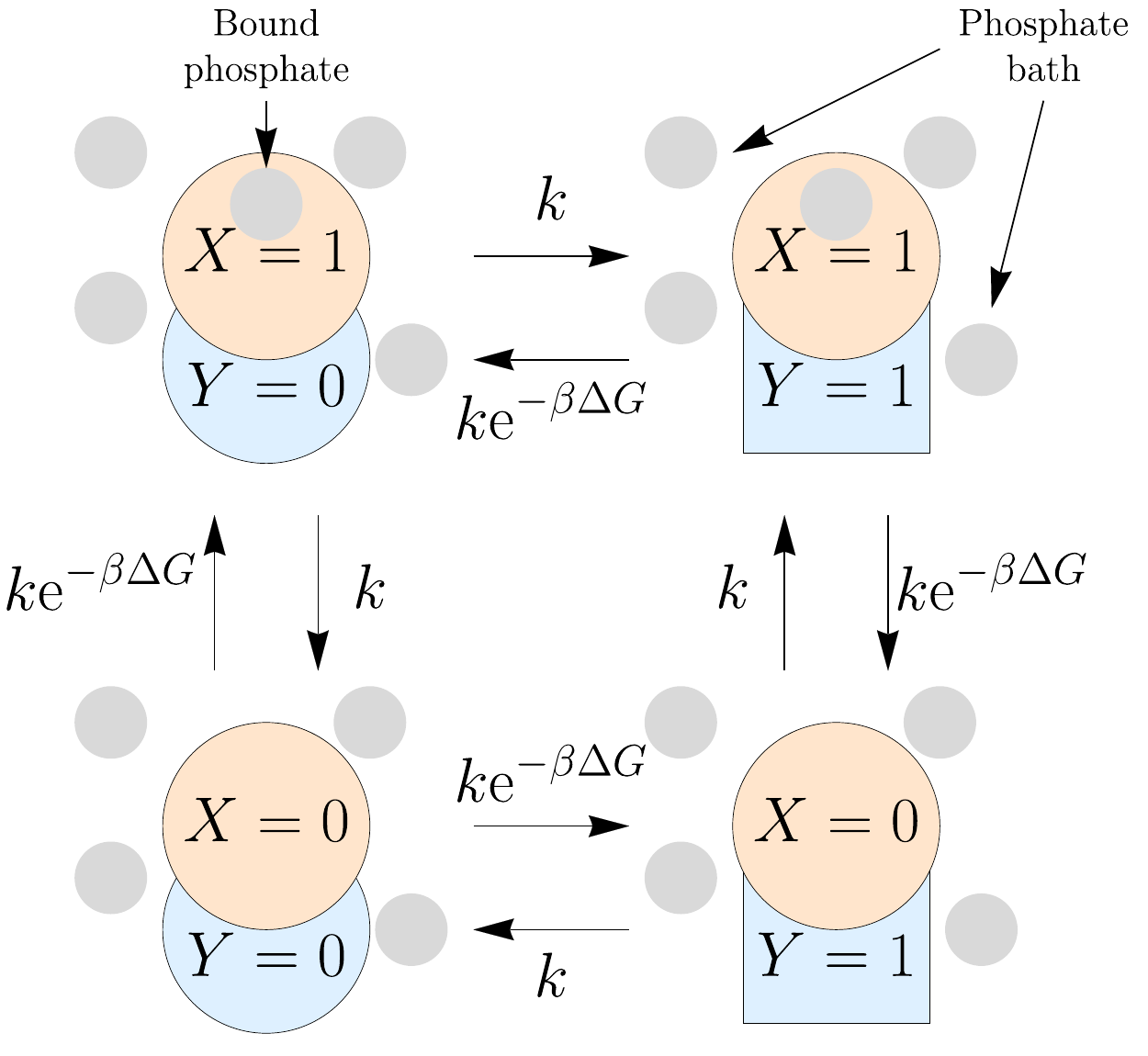}
	\caption{A simple equilibrium molecular system, in which the $X$ and $Y$ represent the configurations of two proteins, X and Y. X can be phosphate-bound or unbound, and Y has two allosteric conformations. Relative rates of forwards and backwards transitions are set by overall free-energy differences.}
	\label{fig:equlibriumgraph}
\end{figure}
The simplest way to violate the assumptions of Section~\ref{sec:system1} is to allow the signal transition rates $w^{xx^\prime}_y$ to depend on $Y,$ whilst retaining the $2\times2$ structure of the overall system. To provide an intuition for such a setting, it is useful to consider the molecular system illustrated in Figure~\ref{fig:equlibriumgraph}, containing two bound molecules, X and Y. Both molecules have two states; for concreteness, we imagine that X can be phosphorylated  (bound to an inorganic phosphate moiety) or unphosphorylated, corresponding to states of a random variable $X= 1$ and $X=0,$ respectively.  Molecule Y, by contrast, can adopt two allosteric configurations, and an energetic coupling between X and Y favours states $X,Y =1,1$ and $0,0$ relative to $1,0$ and $0,1.$ 

If these molecules are in contact only with a bath of phosphate, they will eventually reach a thermodynamic equilibrium described by a Gibbs distribution. Assuming for simplicity that the free energies of the 11 and 00 states are equal, and $\dg$ below the free energies of 10 and 01,
\begin{align}
	p_{10}&=p_{01}=\frac{\mathrm{e}^{-\beta \dg}}{Z}\nonumber\\
	p_{00}&=p_{11}=\frac{1}{Z},
\end{align}
where $Z=2+2\mathrm{e}^{-\beta \dg}$ is the partition function, and $\beta$ is the inverse temperature $1/k_{\mathrm{B}}T.$ At equilibrium, the system must obey detailed balance:
\begin{equation}
	\frac{w^0_{01}}{w^0_{10}}=\frac{w^1_{10}}{w^1_{01}}=\frac{w^{01}_0}{w^{10}_0}=\frac{w^{10}_1}{w^{01}_1}=\mathrm{e}^{-\beta \dg}.
\end{equation}
The simplest choice consistent with this requirement is $w^0_{01} = w^1_{10} =  w^{01}_0 = w^{10}_1 = k\exp(-\beta \dg)$; $w^0_{10} = w^1_{01} =  w^{10}_0 = w^{01}_1 = k.$

The transitions in $X$ are thus strongly influenced by the state of $Y$ --- a natural and necessary feature in an equilibrium system involving coupling between $X$ and $Y$ \cite{Govern_PRL_2014}. One can still evaluate Equation~\ref{eq:other_defn} in this setting, obtaining a nominal $\ly=0$ for all parameter choices. Indeed, $\ly$ is always zero in thermodynamic equilibrium; the contributions arising from each transition in Equation~\ref{eq:other_defn} must cancel with those arising from the reverse transitions. Similarly, Equation~\ref{eq:ly} still holds, but the flux $\phi$ is necessarily zero due to detailed balance, even though $I[X;Y]$ and $\mathrm{Cov}(X,Y)$ can be large ($I[X;Y] = 0.56$ for $\beta \dg = 3.5$).

This equilibrium system and the driven system considered in Section~\ref{sec:system1} are  at opposite ends of a spectrum. In the equilibrium case, the influence of $X$ on $Y$ and $Y$ on $X$ are symmetric, in the sense that both experience the same biasing due to the interaction. Thus, there is no real sense in which Y is a sensor for X; it causes changes in X as much as it reacts to them. For the driven system in Section~\ref{sec:system1}, the influence is totally asymmetric; $X$ influences $Y$ but $Y$ does not influence $X$ at all, which is possible because the external process $X$ is strongly driven. In this setting Y is a purely reactionary sensor.

One might argue that the ``learning rate" is no longer a meaningful term for the quantity defined by the second line of Equation~\ref{eq:informationflows} in the setting where $Y$ influences $X$. However, $\ly$ still quantifies the degree to which $Y$ systematically reacts to (or learns about) $X$. To see this in more detail, we now take the unusual step of interpolating between the two limits of equilibrium and strong driving. We do this  by adding a finite-strength non-equilibrium driving term to the equilibrium system, to explore the response of the learning rate.

\begin{figure}
	\centering
	\includegraphics[width=1\linewidth]{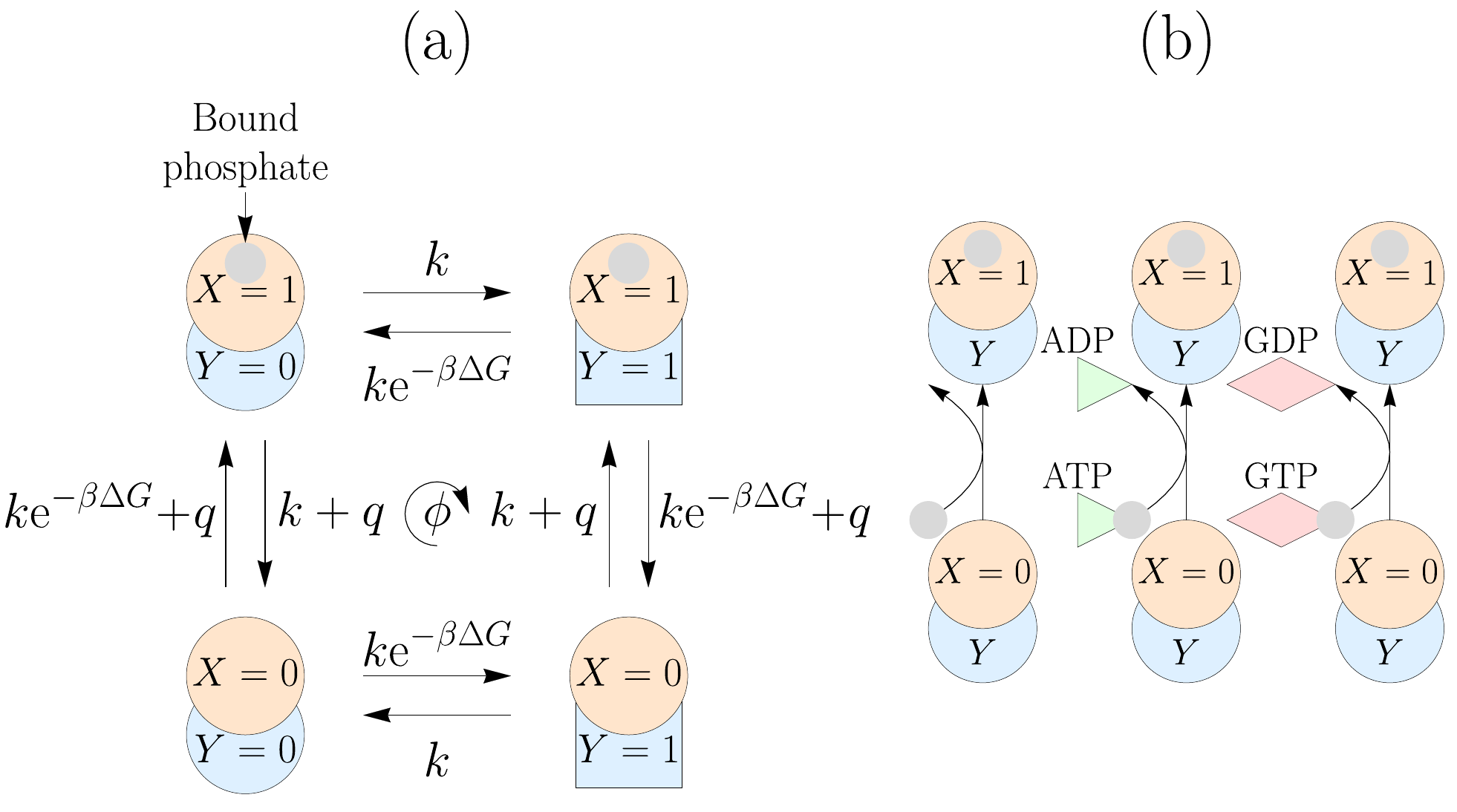}
	\caption{(a) An extension of the system in Figure~\ref{fig:equlibriumgraph} in which it is possible to interpolate between the equilibrium limit and the limit where $X$ is not influenced by $Y$ by changing the ratio $q/k.$ (b) This system could be implemented using three separate pathways for phosphorylation of X. The pathways are by binding phosphate from solution, or via donation of phosphate from ATP or GTP.}
	\label{fig:driving}
\end{figure}

This interpolation takes the form of additional transitions of rate $q$ between X and X$^*,$ independent of the state of molecule Y, alongside the original transitions of rate $k$ and $k\exp(-\beta \dg)$. This system is illustrated in Figure~\ref{fig:driving}\,(a). We can vary $q/k$ to  interpolate between equilibrium ($q/k \rightarrow 0$) and strongly-driven ($q/k \rightarrow \infty)$ systems; in the second case, $X$ is independent of $Y.$

A possible way to implement this scheme in a molecular setting is to drive the system by allowing phosphorylation of X to occur by two additional pathways as well as exchange of phosphate with solution. These additional pathways proceed via exchange of phosphate with nucleotides supplied by  large buffers or chemostats, as shown in Figure~\ref{fig:driving}\,(b). With phosphate P and nucleotides ATP, ADP, GTP and GDP in solution, molecule X can change phosphorylation state by the following reactions
\begin{align}
	\text{P}+\text{X}
	& \rightleftharpoons
	\text{X}^*,\nonumber\\
	\text{ATP}+\text{X}
	& \rightleftharpoons
	\text{ADP}+\text{X}^*,\nonumber\\
	\text{GTP}+\text{X}
	& \rightleftharpoons
	\text{GDP}+\text{X}^*.
\end{align}
Here, X$^*$ represents the molecule in the phosphorylated state. The concentrations of ATP, ADP, GTP and GDP contribute to the overall free energy changes of the second and third reactions; setting their concentrations to be incommensurate with the free energy change for $\text{P}+\text{X} \rightleftharpoons \text{X}^*$ amounts to driving the system so that it cannot reach equilibrium. For example, if [ADP] and [GTP] are both kept low, X will frequently be converted into X$^*$ through interactions with ATP, but will be converted from X$^*$ to X by interactions with GDP. The relative rate of these two reactions is not constrained by the relative rate of  the two reactions involved in direct exchange of phosphate with solution, $\text{P}+\text{X}\rightleftharpoons \text{X}^*,$ since the microscopic processes are distinct. Via this scheme it is therefore possible, at least in principle, to modify the stochastic process as shown in Figure~\ref{fig:driving}\,(a). In particular, we could imagine that ATP/ADP coupling is only possible in the $Y=0$ state, and GTP/GDP coupling is only possible when $Y=1$. The concentrations could then be set so that  the free energy changes of ATP/ADP and GTP/GDP interconversion exactly cancel with the $\pm \Delta G$ associated with the change of state of $X,Y$ in these two cases, allowing phosphorylation/dephosphorylation by these pathways to have equal transition rates.

\begin{figure}
	\centering
	\includegraphics[width=1\linewidth]{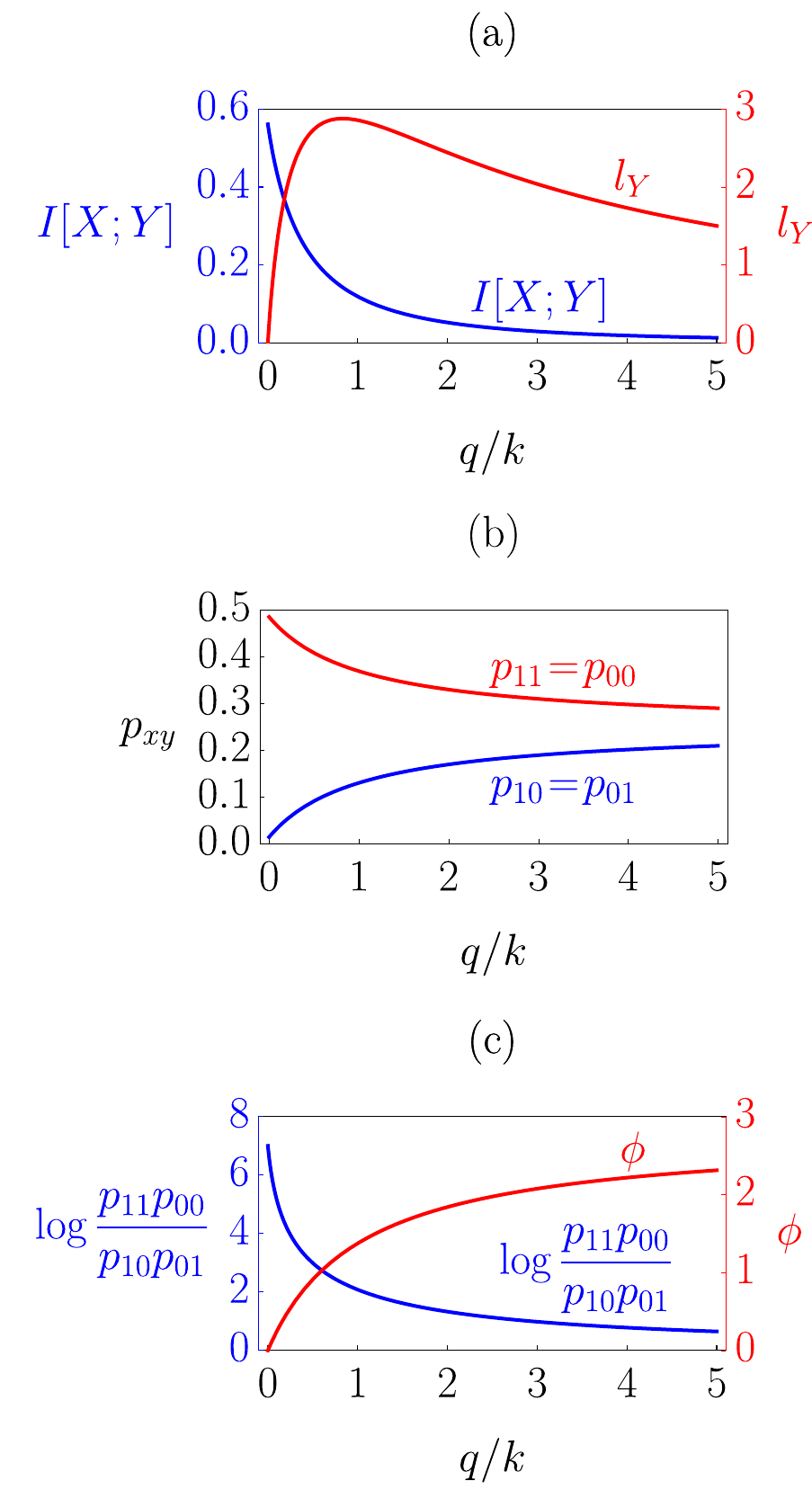}
	\caption{The behaviour of a $2 \times 2$ system that interpolates between equilibrium and driven limits,  illustrated in Figure~\ref{fig:driving}\,(a). Plots obtained with $\beta \dg = 3.5$ and $k=11.5$. (a) The mutual information and learning rate behave distinctly. The learning rate is zero in the equilibrium limit ($q/k \rightarrow 0$) and then increases to a peak with increased driving $q.$ The mutual information is highest at equilibrium and decreases as driving increases. (b) The non-equilibrium steady state probabilities. (c) Two competing contributions to $\ly$; the flux, $\phi$, and information affinity, $\ln (p_{11}p_{00}/p_{10}p_{01}),$ show opposite dependencies on $q/k,$ leading to a peak at finite $q/k.$}
	\label{fig:feedbackplot}
\end{figure}

We perform the interpolation between equilibrium and driven systems in Figure~\ref{fig:feedbackplot}\,(a), where we have plotted the mutual information and learning rate against $q/k$ for $\beta \dg=3.5$ at steady state. In this system the mutual information and learning rate behave quite differently. At $q/k=0,$ $I[X;Y]$ is large and its size is set by the energy gap; as $\beta \dg \rightarrow \infty,$ $I[X;Y] \rightarrow \ln2,$ corresponding to perfect correlation. As $q/k$ increases,  $X$ transitions start to become increasingly decoupled from the state of $Y,$ and also much faster. Correlations are thus destroyed as shown in Figure~\ref{fig:feedbackplot}\,(b). The $X$ state changes randomly and too fast for $Y$ to track it and consequently,  $I[X;Y] \rightarrow 0.$

By contrast, the learning rate is zero for $q/k=0,$ has a peak at $q/k \sim 1$ and decays to zero as $q/k\rightarrow \infty.$ To understand this behaviour, it is helpful to consider Equation~\ref{eq:ly}. For a $2\times2$ state system the learning rate is equal to the flux $\phi$ around the states multiplied by the information affinity. As shown in Figure~\ref{fig:feedbackplot}\,(c), the information affinity behaves very similarly to $I[X;Y].$ This is because, as we have discussed in Section~\ref{sec:system1}, the information affinity measures the correlation between $X$ and $Y$ and increasing $q/k$ decreases the correlation. Therefore, the ratio $\frac{p_{11}p_{00}}{p_{01}p_{10}}$ becomes closer to 1 so the information affinity decreases.

In contrast to the information affinity, the flux $\phi$ increases with increased driving strength $q/k.$ The reason for this is that as $q/k \rightarrow 0$, the system tends towards a detailed balanced equilibrium state, whereas as $q/k$ increases the system is increasingly dominated by the non-equilibrium drive, which causes a current. The combination of the opposing  behaviours of the information affinity and the flux is the peaked shape of $\ly.$   It may seem surprising that increasing $q/k$ causes opposing behaviour of the information affinity and its conjugate current $\phi$. However, in general affinities depend on relative (not absolute) rates, whereas fluxes are determined by absolute rates --- similar behaviour therefore arises in general systems.   In any case, now that the assumption of no feedback from $Y$ to $X$ is violated, it is no longer true that $\phi$  and hence $\ly$ reflect the strength of correlation, as in Section~\ref{sec:system1}.

What intuition, then, do we gain from the learning rate? The presence of a flux $\phi>0$ is indicative of the fact that $Y$ is responding to changes in $X$; systematically, $X$ tends to transition from 0 to 1 before $Y,$ which then tends to follow. The dependence of $\ly$ on $\phi$ thus reflects the fact that the learning rate measures the degree to which transitions in $Y$ respond to transitions in $X$ to maintain correlations, rather than the strength of correlations themselves. This is apparent from the definitions in Equations~\ref{eq:dIdt} and \ref{eq:nostalgia}. In a system with feedback from $Y$ to $X,$ these two metrics report on quite distinct properties. In the equilibrium limit, correlations are strong but transitions in $Y$ are just as likely to precede transitions in $X$ as to follow them, and hence the learning rate is zero.

\subsection{A more complex signal process}
\label{sec:system3}
In Section~\ref{sec:system2}, we relaxed the assumption that the downstream system $Y$ did not influence the upstream system $X.$ As a result, the tight connection between learning rate and correlation between $X$ and $Y,$ observed in Section~\ref{sec:system1}, was broken. We now explore the possibility of introducing more complex upstream signals $X$ than assumed in Section~\ref{sec:system1}, whilst restoring the assumption that $X$ is not influenced by $Y.$ We still consider systems in the stationary state.

\begin{figure}
	\centering
	\includegraphics[width=1\linewidth]{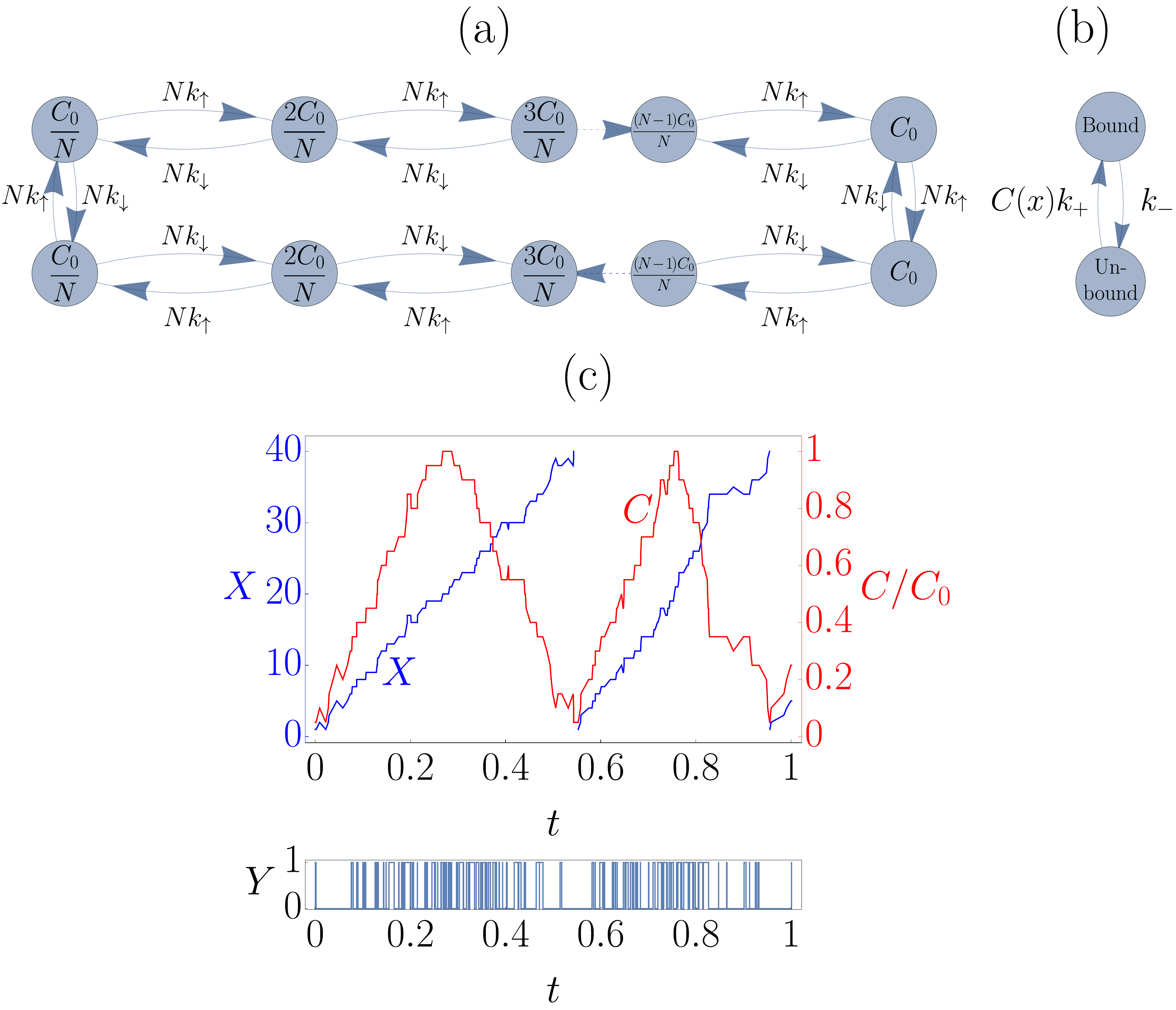}
	\caption{A bipartite system in which a receptor modelled by $Y$ responds to an oscillating concentration determined by $X.$ (a) A graphical representation of the process $X,$ indicating the ligand concentration $C$ for each $X.$  $X$ is biased to move clockwise around the loop so $C$ is driven up and then down in an oscillating fashion. $X$ is not influenced by $Y.$ (b) For $Y,$ the binding rate is proportional to the ligand concentration and the unbinding rate is constant, as in Section~\ref{sec:system1}. (c) Typical behaviour of $X,$ $C$ and $Y$ over time (for $N=20,$ $C_0=1,$ $\ku=5,$ $\kd=0.5$ and	 $\kp=\km=300$), showing stochastic oscillations in concentrations}
	\label{fig:cyclicgraph}
\end{figure}

As a physical model to provide intuition, we again consider a model of concentration sensing by a single  receptor. In this case, we consider an oscillating concentration signal \cite{Becker2015}, as might be relevant to a cell experiencing roughly periodic, but noisy, fluctuations in its environment. Such a situation might be relevant to cells in animal intestines, for example. To construct an oscillating concentration, we consider a Markov process, $X$, with $2N$ discrete states, as shown in Figure~\ref{fig:cyclicgraph}\,(a). States of $X$ with $x=1...N$ have a concentration $C= xC_0/N,$ and states $x=N+1...2N$ have a concentration $C= (2N+1-x)C_0/N.$ Thus there are $N$ distinct concentrations, each present at two values of $X.$ We consider transitions between only neighbouring $X$ states, with $w^{x\,x+1}=N\ku$ and  $w^{x\,x-1}=N\kd.$ 

 This partitioning of $X$ into two ``rows" allows the construction of a Markov process $X$ that leads to noisy oscillations in the concentration $C,$ between $\tfrac{C_0}{N}$ and $C_0.$ Taking $N\ku > N\kd,$ the system will tend to move to higher values of $C$ when $1 \leq x \leq N,$ and towards lower values when $N+1 \leq x \leq 2N$ (clockwise in Figure~\ref{fig:cyclicgraph}\,(a)). Scaling rate constants with $N$ means that the overall flux around the cycle is $N$-independent. The typical behaviour of $X(t)$ and $C(t)$ is shown in Figure~\ref{fig:cyclicgraph}\,(c). A similar Markov chain biased by a constant chemical potential to produce noisy oscillations was used by Barato and Seifert \cite{Barato_Cost_2016} to model a clock.

\begin{figure}
	\centering
	\includegraphics[width=1\linewidth]{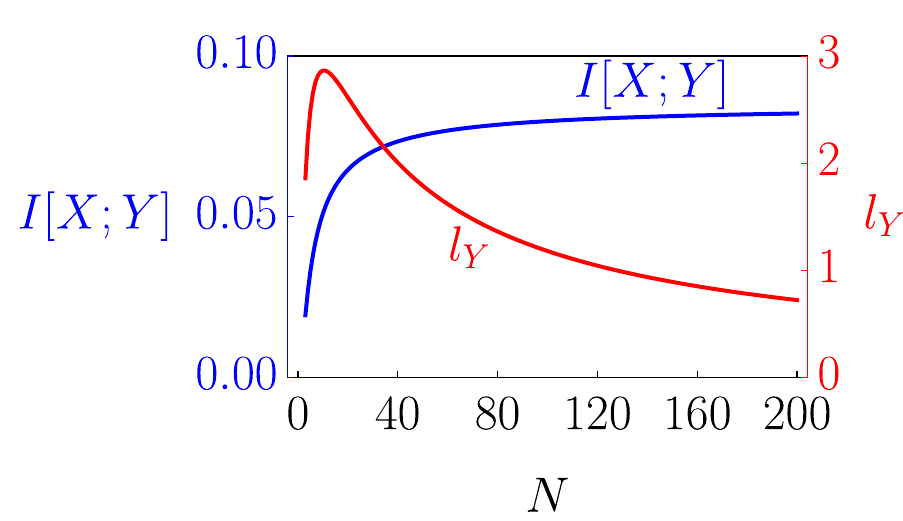}
	\caption{The learning rate and instantaneous mutual information show opposite behaviours as the number of external states $N$ is varied for the cyclic system illustrated in Figure~\ref{fig:cyclicgraph}\,(a). The learning rate decays from a finite value to zero, whereas the instantaneous mutual information increases monotonically to a finite plateau. Data obtained for $\ku=20,$ $\kd=10,$ $\kp=10000,$ $\km=2000$ and $C_0=1.$}
	\label{fig:cyclicplot}
\end{figure}

The receptor state defines the sub-process $Y;$ states and transitions are shown in Figure~\ref{fig:cyclicgraph}\,(b). The receptor properties are identical to that considered for the simpler system in Section~\ref{sec:system1}.  There is a constant rate of unbinding, $w^{x}_{10}=\km,$ and the binding rate $w^{x}_{01}=C(x)\kp$ is proportional to the ligand concentration.

From the perspective of the learning rate, it is illuminating to vary $N$ (the number of discrete concentration states) whilst fixing all other parameters. As is evident from Figure~\ref{fig:cyclicplot}, $I[X;Y]$ increases with $N$ up to a finite plateau, whereas the learning rate decreases to zero. The behaviour of $I[X,Y]$ is intuitive; as more states are added, it becomes easier to reliably distinguish values of $X$ that correspond to high and low values of $C.$ In particular, for low $N,$ $C$ changes in large jumps due to $X$ transitions, meaning that $Y$ is an inaccurate reporter for $X$ immediately after the transition. For larger $N,$ the change in $C$ is smoother. Eventually, however, this effect saturates, and $I[X;Y]$ reaches a finite plateau.

By contrast, the learning rate eventually falls with large $N,$ tending to zero in the limit $N\rightarrow \infty$ (we present a general proof in Appendix~\ref{app:derivation}). To understand this difference in behaviour, note that our choice of  $w^{x\,x+1}=N\ku$ and  $w^{x\,x-1}=N\kd$ ensures that the overall average rate at which the concentration undergoes oscillations is fixed, regardless of $N.$ The number of forward and backward steps of the process $X$ in a time $\Delta t$ are independently Poisson-distributed with means $N\ku \Delta t$ and $N\kd \Delta t,$ respectively. Thus, the mean net number of  forward steps in $\Delta t$ is $\Delta t N(\ku-\kd),$ proportional to $N,$ compensating for the fact there are more states covering the same concentration window. The standard deviation in the net number of forward steps in $\Delta t$ is $\sqrt{\Delta t N(\ku+\kd)},$ however. Consequently, as $N$ is increased, the relative uncertainty in the net number of forward steps during $\Delta t$ falls as $1/\sqrt{N}.$ Larger $N$ therefore implies a smaller relative error in predicting the future value of $X,$ given knowledge of its current value. Increasing the number of states $N$ is thus effectively a method to interpolate between highly stochastic and quasi-deterministic behaviour of $X$. Note that it is the convergence on deterministic behaviour of the signal that in general takes $\ly\rightarrow 0$; not $N \rightarrow \infty.$

As $N \rightarrow \infty,$ therefore, the rate at which the current value of $Y$ becomes ineffective in predicting future values of $X$ tends towards zero. 
In other words, the ``nostalgia" of $Y$ (Equation~\ref{eq:nostalgia} and Reference~\cite{still_thermodynamics_2012}), and hence $\ly,$ tends towards zero. The learning rate quantifies the rate  at which transitions in $Y$ must respond to transitions in $X$ to maintain a steady-state information; this rate is zero for large $N$ despite $I[X;Y]$ increasing monotonically with $N.$ Thus, using a more complex signal than in the simple $2\times2$ network in Section~\ref{sec:system1} also serves to highlight the distinctions between the physical meaning of $\ly$ and $I[X;Y].$

\subsection{Optimisation of sensors}
Due to its provenance as an informational quantity, $\ly$ has been proposed as a metric for sensor performance in steady-state sensing circuits \cite{barato_efficiency_2014, hartich_sensory_2016}. In Sections~\ref{sec:system2} and \ref{sec:system3}, however, we have emphasised that $\ly$ reflects the rate at which transitions in the downstream subsystem $Y$ respond to changes in $X$ to maintain a steady-state information $I[X;Y].$ There is no reason {\it a priori} to suppose that maximising such a quantity is inherently optimal for a sensor --- indeed, one might imagine that having to compensate for transitions in $X$ at a lower rate would be preferable in reliable sensing. Further, our results in Sections~\ref{sec:system2} and \ref{sec:system3}  show that the response of $\ly$ to parameter variation can be unrepresentative of the degree to which $Y$ relates to signal $X.$   

Sections~\ref{sec:system2} and \ref{sec:system3}, however, cannot reasonably be described as sensor optimisation. To optimise a sensor, it is natural to consider a setting in which the dynamics of signal $X$ are fixed, and optimisation is performed over the parameters of the downstream network $Y.$ In Sections~\ref{sec:system2} and \ref{sec:system3}, however, we considered the response of $\ly$ to variations in the dynamics of $X.$ Moreover, in Section~\ref{sec:system2}, we even considered  a ``downstream" network $Y$ which influences the ``upstream" network $X.$ This is hardly a well-defined sensor. In the simple system of Section~\ref{sec:system1}, for which we did consider variation over the parameters of the $Y$ sub-process, and $X$ was not influenced by $Y,$ optimising $\ly$ and $I[X;Y]$ over sensor parameters was essentially equivalent. It is therefore worth considering whether signal/sensor architectures do exist in which optimising over the sensor parameters for $\ly$ and $I[X;Y]$ give markedly different results.

\begin{figure}
	\centering
	\includegraphics[width=\linewidth]{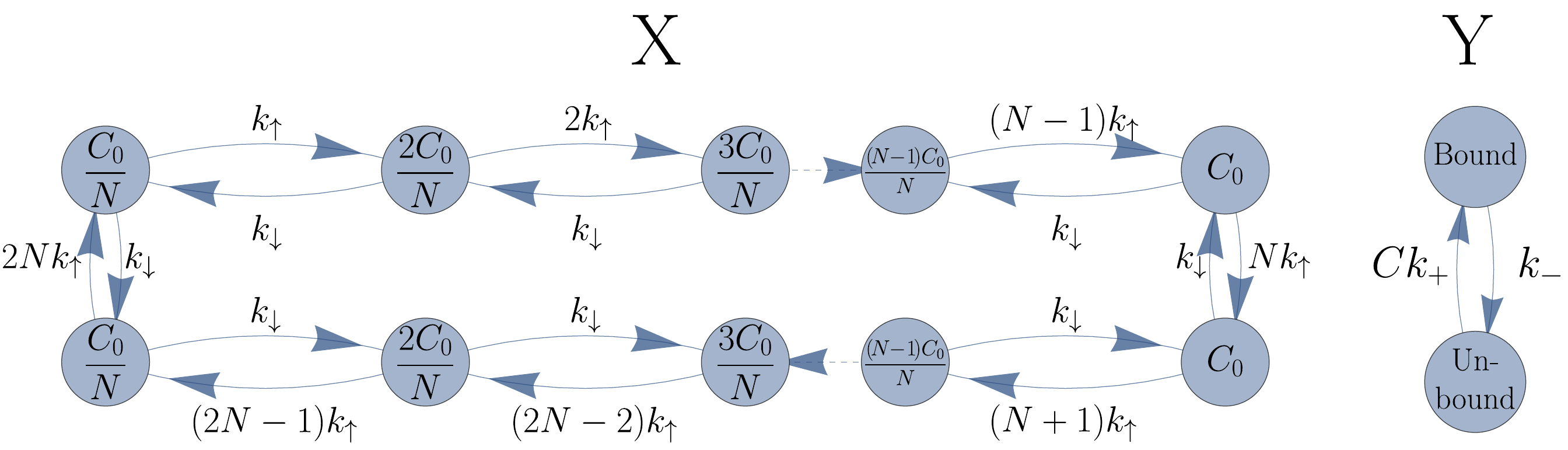}
	\caption{A bipartite system $X,Y$ for which optimising the parameters of the downstream subsystem $Y$ to maximise learning rate and mutual information give markedly different results. The system is essentially identical to that considered in Figure~\ref{fig:cyclicgraph}, but with hopping rates for the upstream process $X$ that depend on $x.$}
	\label{fig:xasymmetricgraph}
\end{figure}

To demonstrate this possibility, we consider a system identical to the oscillating concentration sensor considered in Section~\ref{sec:system3}, but for which the transition rates of the $X$ sub-process are $x$-dependent (Figure~\ref{fig:xasymmetricgraph}). The sensor $Y,$ with transition rates $w^{x}_{10}=\km$ and $w^{x}_{01}=C(x)\kp,$ is unchanged.  

\begin{figure}
	\centering
	\includegraphics[width=\linewidth]{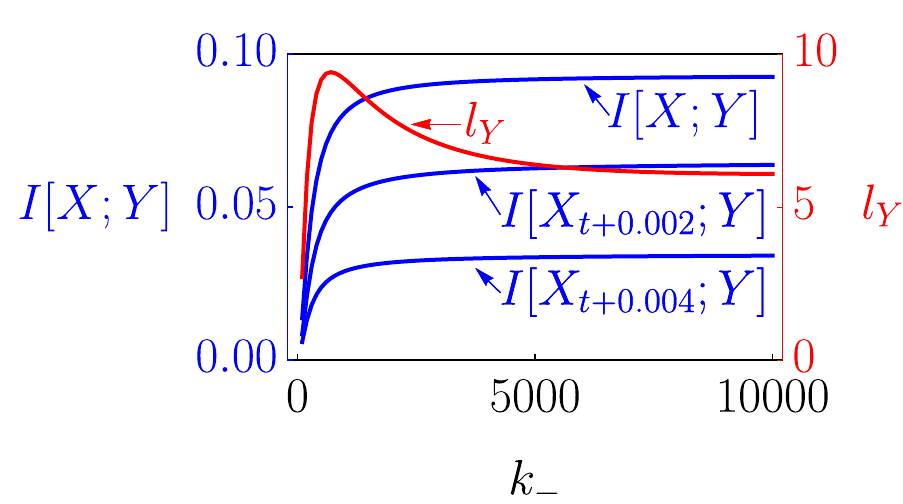}
	\caption{For a sensor detecting an oscillating system with a non-uniform hopping rate, $I[X;Y]$ increases monotonically with the relative response rate of the sensor, whereas $\ly$ shows a peak at a finite response rate. The two metrics thus imply distinct ``optimal" sensors. Also shown are some examples of $I[X_{t+\tau};Y_t],$ where $\tau$ is a positive time delay. These too are monotonic in response rate. Data obtained for $N=20,$ $\ku=200,$ $\kd=10,$ $\kp/\km=2$ and $C_0=1$}
	\label{fig:asymmetricplot}
\end{figure}

We consider optimising over the relative rates of the $X$ and $Y$ subsystems at  fixed  number of $X$ states $N,$ fixed transition rates in the $X$ subsystem, and fixed $\kp/\km$ within $Y.$ In Figure~\ref{fig:asymmetricplot}, we show that the instantaneous mutual information  increases monotonically with the rates of the downstream subsystem (represented by $\km$), from $I[X;Y]=0$ up to a plateau. By contrast, the learning rate increases from $\ly=0$ to a peak at finite $\km,$ before decaying to a plateau. Maximising the instantaneous mutual information and maximising the learning rate give qualitatively different ``optimal sensors".

We have considered optimising over this subspace with $\kp/\km$ fixed rather than the whole $(\kp,\km)$ space for ease of illustration. For a metric to be useful for sensor optimisation it must be able to meaningfully differentiate sensing quality between any two sets of parameters, and provide a useful optimisation within any subspace of the parameter space. We note that similar behaviour is observed for other values of $\kp/\km$.

The ``optimal sensor" indicated by the instantaneous mutual information is intuitively reasonable. When the rates of the sensor are as large as possible, $Y$ can keep up with $X$; otherwise it simply lags behind, reporting earlier values of $X.$ Whilst it is possible that a slower-responding sensor could provide lower $I[X,Y],$ but increased predictive power at some specific future time $\tau,$  $I[X_{t+\tau};Y_t]$ \cite{Becker2015}, we find no evidence of this for our case (see Figure~\ref{fig:asymmetricplot}). Therefore, the instantaneous mutual information provides a metric for sensor performance that is intuitively reasonable, if not necessarily unique. By contrast, we can find no intuitive justification of the peak of the learning rate in terms of optimal sensing.  Indeed, the peak in the learning rate  is reminiscent of the peak in the dissipation rate $\sigma_Y$ that arises in such systems \cite{Becker2013}. This peak is indicative of a subsystem $Y$ that lags behind $X$ sufficiently that the response to driving is highly irreversible, but is not so slow that it barely responds at all \cite{Becker2013}. Similarly, it is plausible that the peak in $\ly$ reflects a delay in $Y$ relative to  $X$ that its large enough to  ensure that transitions have a strong tendency to increase $I[X;Y],$ but not so large that $Y$ barely responds to changes in $X$ at all. Regardless, since we are unable to explain the location of this peak through any optimal tracking perspective,  it is difficult to see how $\ly$ is informative of quality of sensing.

\section{Conclusions}
We have explored the physical interpretation of the learning rate $\ly,$ a recently proposed statistical metric defined on bipartite  stochastic systems \cite{barato_efficiency_2014, hartich_sensory_2016,zhang_critical_2016}.  $\ly$ quantifies the rate at which transitions in a downstream system $Y$ act to increase the mutual information between $Y$ and an upstream system $X,$ or the rate at which future values of $Y$ become more predictive of the current values of $X$ than the current value of $Y$. In the steady state, a biophysically relevant scenario, $\ly$ is also equivalent to the rate at which transitions in $X$ reduce the mutual information, or the ``nostalgia" rate at which information about the current value of $X$ becomes irrelevant. 

We have shown that, in the simplest steady-state bipartite system in which $X$ is not influenced by $Y,$ $\ly$ essentially reports on the correlation between subsystems. In other settings $\ly$ can show quantitatively and qualitatively different behaviour from measures of interdependence such as the mutual information and covariance. Moreover, we have demonstrated that this difference can be explained by the fact that $\ly$ quantifies the rate at which transitions in $Y$ act to maintain a steady state $I[X;Y],$ rather than the  magnitude of $I[X;Y].$ In general, there is no reason why these two properties should behave similarly.

Fundamentally, since $l_Y$ represents the rate at which transitions in $Y$ act to increase information between $X$ and $Y$, the ``learning rate" is a reasonable description of this quantity. However, the rate of learning and quality of sensing are not directly related, and we do not find that $\ly$ is a good general metric for sensor performance, as has been proposed \cite{barato_efficiency_2014, hartich_sensory_2016}.  We have shown that, in at least one signal/sensor context, maximising the learning rate over a subset of  sensor parameters gives a result that is hard to interpret as an optimal sensor. Without first identifying how the learning rate is reporting on sensor performance in a given context, it is difficult to see how $\ly$ can be reliably used as a metric for sensing  in general.

 One can imagine two sensors, $Y$ and $Y^\prime,$ of the same signal $X,$ which produce very different  joint probability distributions $p(x,y)$ and $p(x,y^\prime)$ but the same instantaneous mutual information $I(X,Y) = I(X,Y^\prime)$. It is perfectly possible, for example, that $Y$ predicts the future of $X$ less well than $Y^\prime$, despite the instantaneous mutual information being equal, if the fluctuations in $X$ coupled to by $Y$ are less informative about system dynamics than those reported by $Y^\prime$. In this case, to maintain equal information in the steady state, $Y$ will need to refresh its information more rapidly, and hence we would have $l_Y > l_{Y^\prime}$. However, it would seem unreasonable to classify $Y$ as a better sensor than $Y^\prime$; indeed, if anything, $Y^\prime$ seems more effective since it possesses the same information about the present state of $X$ and more information about its future evolution. High $l_Y$ may therefore be indicative of a sensor that is in some sense inefficient: constantly needing to update information that rapidly becomes worthless.  

If $l_Y$ is not a direct measure of sensor performance, the fact that $l_Y$ is bounded  by the entropy production due to $Y$-transitions, $\sigma_Y > l_Y$, does not imply that entropy production is limiting for sensing. However, we do not argue that entropy production is not limiting for sensing --- only that such a justification via the learning rate would be flawed. As argued by Govern and ten Wolde \cite{Govern_PRL_2014}, in any equilibrium system, there is no real distinction between ``signal'' and ``readout'' --- both must necessarily influence each other, and therefore the construction of a true sensor is impossible. In Reference~\cite{Das_A_2016} a bound relating entropy generation to instantaneous mutual information is derived in the case where the sensor does not influence the signal.

Our work has considered only a small set of possible Markovian systems, and has not touched on issues such as performance of readout networks that  are designed to integrate a signal over time \cite{govern2014,ouldridge_thermodynamics_2017} or perform more complex operations such as differentiation \cite{Becker2015}. It is possible that the learning rate reflects an  aspect of the performance in these circuits. Further, we have not investigated the behaviour of the learning rate in systems that are yet to reach steady state. In such cases, $\ly$ is no longer equal to either the nostalgia rate, or the rate at which transitions in $X$ tend to reduce $I[X;Y].$ Moreover, it is potentially non-zero even for passive (non-driven) systems. The meaning of the learning rate in such cases warrants further consideration.

Finally, the learning rate can provide insight into the functioning of systems even if it is not a metric for performance of sensing or similar functionality. For example, its properties were recently applied to derive results that do not depend on its interpretation in a simple model of a perceptron \cite{Gold_Stochastic_2017}. $\ly$ can also reasonably be interpreted as an indicator of the direction and magnitude of information flow in a steady-state bipartite network, as evidenced by its role in the refined second law (Equation~\ref{eq:bipartite_second_law}). In the context of a model of an autonomous Maxwell's demon it is the flow of information between the `demon' subsystem and the `engine' subsystem \cite{horowitz_thermodynamics_2014}.

One interpretation of the learning rate in steady state might be as a quantification of the effort put in by the downstream system to maintain a certain interdependence of $X$ and $Y,$ regardless of whether this interdependence is optimal in some sense.  For example, if the magnitude of all rates is increased but their ratios kept constant, then the learning rate is increased although there is no change in the correlation between the sensor and signal. Indeed, this sense of $l_Y$ as a measure of how hard the downstream system has to work is apparently reinforced by the relation $ l_Y< \sigma_Y$, with $\sigma_Y$ being the entropy generation due to transitions in the Y-subsystem. It should be noted, however, that although entropy is generated by $Y$ transitions, this is not necessarily indicative of a consumption of resources held by the $Y$ subsystem \cite{barato_efficiency_2014}. In the biomolecular examples given here, for example, the $Y$ subsystems are totally passive, consuming no chemical fuel. The thermodynamic work is done by the driving of the external process $X$ \cite{barato_efficiency_2014}, which is particularly evident in the example of Section~\ref{sec:system2}, where the nucleotides fuelling the driving are explicitly considered. From the perspective of a sensor or a cell, entropy generation arising from  environmental changes, rather than due to internal fuel consumption, is not obviously a limiting factor.
 
\section{Acknowledgements}
We acknowledge Susanne Still for a discussion of the topics and Pieter Rein ten Wolde, Andre Barato, Udo Seifert, David Hartich and Sebastian Goldt for commenting on the manuscript.

T. E. O. acknowledges support from a Royal Society University Research Fellowship and R. A. B. acknowledges support from an Imperial College London AMMP studentship.

\bibliography{papers}
\bibliographystyle{mybibstyle}

\begin{appendix}
\begin{widetext}
\section{Proof that $\ly$ tends to zero as $N \rightarrow \infty$ for the system in Section~\ref{sec:system3}}
\label{app:derivation}
Using Equations~\ref{eq:informationflows} and \ref{eq:steadystate}, in the steady state
\begin{equation}
	\ly=-\dot{I}^X=-\sum_{xx'y}w^{xx'}p(x,y)\log\frac{p(y|x')}{p(y|x)}\nonumber=\sum_{xx'y}w^{xx'}p(x,y)\big(\log p(y|x)-\log p(y|x')\big).
\end{equation}
Relabelling dummy variables,
\begin{equation}
	\ly=\sum_{xx'y}\left(p(x,y)w^{xx'}-p(x',y)w^{x'x}\right)\log p(y|x).
\end{equation}
In this system, $p(x)=\tfrac{1}{2N}$ for all $x$ by symmetry. There are only transitions between adjacent $X$ states and so

\begin{equation}
	\ly(N)=\frac{1}{2N}\sum_{xy}\left(p_N(y|x)w^{xx^+}-p_N(y|x^+)w^{x^+x}+p_N(y|x)w^{xx^-}-p_N(y|x^-)w^{x^-x}\right)\log p_N(y|x)
\end{equation}
where $x^+$ is the next $X$ state in the cycle and $x^-$ is the previous state, and we have introduced an explicit dependence of probabilities on $N.$ In this system
\begin{align}
	w^{xx^+}=w^{x^-x}=N\ku,\nonumber\\
	w^{x^+x}=w^{xx^-}=N\kd,
\end{align}
and therefore
\begin{equation}
	\ly(N)=\frac{1}{2}\sum_{xy}\Big(\ku\big(p_N(y|x)-p_N(y|x^-)\big)+\kd\big(p_N(y|x)-p_N(y|x^+)\big)\Big)\log p_N(y|x).
\end{equation}
Defining $z=\tfrac{x}{2N}$ we have
\begin{equation}
	\ly=\frac{1}{2}\sum_{zy}\Big(\ku\big(p_N(y|z)-p_N(y|z-\tfrac{1}{2N})\big)+\kd\big(p_N(y|z)-p_N(y|z+\tfrac{1}{2N})\big)\Big)\log p_N(y|z).
\end{equation}
We now assume that for large $N,$ $p_N(y|z)$ tends smoothly towards $p_\infty(y|z)$, where $p_\infty(y|z)$ is a smooth function of $z$ with $p_\infty(y|0)=p_\infty(y|1)$ due to the boundary conditions that connect $z=1$ ($x=2N$) to $z=0$ ($x=0$). For sufficiently large $N,$
\begin{align}
p_N(y|z) = p_{\infty}(y|z) + \frac{1}{N} \left. {\frac{\mathrm{d}p_N(y|z)}{\mathrm{d}1/N}} \right\rvert_{N \rightarrow \infty} +\mathcal{O}\left(\frac{1}{N^2}\right).
\end{align}
Thus
\begin{equation}
	p_N(y|z\pm\tfrac{1}{2N})=p_\infty(y|z)\pm\frac{1}{2N}\dir{p_\infty(y|z)}{z}+  \frac{1}{N} \left. {\frac{\mathrm{d}p_N(y|z)}{\mathrm{d}1/N}}\right\rvert_{N \rightarrow \infty} +\mathcal{O}\left(\frac{1}{N^2}\right)\nonumber
\end{equation}
and the sum can be approximated by an integral
\begin{equation}
	\sum_z\tfrac{1}{2N}f(z)\rightarrow\int_0^1\mathrm{d}zf(z)+\mathcal{O}\left(\frac{1}{N}\right).
\end{equation}
So for large $N,$
\begin{equation}
	\ly(N)=\frac{1}{2}\sum_y\int_0^1\mathrm{d}z \left( (\ku-\kd)\dir{p_\infty(y|z)}{z} \log p_\infty(y|z)\right)+\mathcal{O}\left(\frac{1}{N} \right).
\end{equation}
Performing the integral within the sum gives
\begin{align}
	\int_0^1\mathrm{d}z\dir{p_\infty(y|z)}{z}\log p_\infty(y|z)&=\Big[p_\infty(y|z)\big(\log p_\infty(y|z)-1)\big)\Big]_0^1\nonumber\\
	&=0,
\end{align}
where the second equality follows from $p_\infty(y|0)=p_\infty(y|1)$. Therefore, $\ly(N) \sim \mathcal{O}\left(\frac{1}{N} \right)$ and thus tends to zero in the limit of $N \rightarrow \infty.$
\end{widetext}
\end{appendix}

\end{document}